\documentclass[aps,prd,superscriptaddress,preprint,tightenlines,nofootinbib]{revtex4}

\usepackage{graphicx}
\usepackage{dcolumn}
\usepackage{bm}

\newcommand{\Ks}{K^0_S}
\newcommand{\Kl}{K^0_L}

\newcommand{\BF}{{\mathcal B}}
\newcommand{\Mbc}{M_{\rm BC}}
\newcommand{\MMsq}{M^2_{\rm miss}}

\begin{document}

\preprint{CLNS 07/2008}
\preprint{CLEO 07-13}

\title{Comparison of $D \to K^0_S \pi$ and $D \to K^0_L \pi$ Decay Rates}

\author{Q.~He}
\author{J.~Insler}
\author{H.~Muramatsu}
\author{C.~S.~Park}
\author{E.~H.~Thorndike}
\author{F.~Yang}
\affiliation{University of Rochester, Rochester, New York 14627}
\author{T.~E.~Coan}
\author{Y.~S.~Gao}
\affiliation{Southern Methodist University, Dallas, Texas 75275}
\author{M.~Artuso}
\author{S.~Blusk}
\author{J.~Butt}
\author{J.~Li}
\author{N.~Menaa}
\author{R.~Mountain}
\author{S.~Nisar}
\author{K.~Randrianarivony}
\author{R.~Sia}
\author{T.~Skwarnicki}
\author{S.~Stone}
\author{J.~C.~Wang}
\author{K.~Zhang}
\affiliation{Syracuse University, Syracuse, New York 13244}
% ...
\collaboration{CLEO Collaboration}
\noaffiliation

\date{November 7, 2007}

\begin{abstract}
We present measurements of $D \to K^0_S \pi$ and $D \to K^0_L \pi$
branching fractions using 281 pb$^{-1}$ of $\psi(3770)$ data at the CLEO-c experiment.
We find that $\BF(D^0 \to \Ks\pi^0)$ is larger than $\BF(D^0 \to \Kl\pi^0)$,
with an asymmetry of $R(D^0) = 0.108 \pm 0.025 \pm 0.024$.  For $\BF(D^+ \to \Ks\pi^+)$
and $\BF(D^+ \to \Kl\pi^+)$, we observe no measurable difference; the asymmetry is
$R(D^+) = 0.022 \pm 0.016 \pm 0.018$.  The $D^0$ asymmetry is consistent with the value based on the U-spin
prediction $A(D^0 \to K^0 \pi^0)/A(D^0 \to \bar K^0 \pi^0) = -\tan^2
\theta_C$, where $\theta_C$ is the Cabibbo angle.
\end{abstract}

\maketitle

As the dominant decay of the charm quark is to the strange quark, the final states of $D^0$ and $D^+$ meson decays typically include $K^+$, $K^-$, $K_S^0$, and $K_L^0$ mesons.  While there have been many measurements of $D$ decays to final states containing $K^\pm$ and $K_S^0$, until now there have been no measurements of decays to final states containing a $K_L^0$.  Typically it has been assumed that the branching fraction for a decay $D \to K_L^0 X$ will equal that for the corresponding decay $D \to K_S^0 X$.  However, as pointed out by Bigi and Yamamoto \cite{bigi}, interference between
Cabibbo-favored transitions (producing an $s$ quark, and thus a $\bar{K}^0$) and doubly-Cabibbo-suppressed transitions (producing an $\bar{s}$ quark, and thus a $K^0$) can lead to a difference in the rates for $D \to K_L^0 X$ and $D \to K_S^0 X$.  Here we present first measurements of two $D$ decays to $K_L^0$, $D^0 \to K_L^0 \pi^0$ and $D^+ \to K_L^0 \pi^+$, and we compare the branching fractions with those for $D^0 \to K_S^0 \pi^0$ and $D^+ \to K_S^0 \pi^+$.  (Throughout, charge-conjugate modes
are implied, except where noted.)  These comparisons provide information about amplitudes and strong phases in $D \to K \pi$ decays.

For these measurements we use a 281 pb$^{-1}$ sample of $e^+e^- \to \psi(3770)$
events, produced by the CESR-c storage ring and recorded with the CLEO-c detector.  The CLEO-c detector is a general purpose solenoidal detector which includes a tracking
system for measuring momentum and specific ionization ($dE/dx$) of charged particles, a Ring
Imaging Cherenkov detector (RICH) to aid in particle identification, and a CsI calorimeter
for detection of electromagnetic showers. The CLEO-c detector is described in detail
elsewhere \cite{bib:cleo_1,bib:cleo_2,bib:cleo_3}.

The $\psi(3770)$ resonance
is below the threshold for $D\bar{D}\pi$, and so the events of interest, $e^+e^-\to
\psi(3770)\to D\bar{D}$, have $D$ mesons with energy equal to the beam energy and a
unique momentum.  Thus, for identifying $D^0$ and $D^+$ candidates, we follow Mark III \cite{bib:MarkIII} and
define the two variables $\Delta E$ and beam-constrained mass $\Mbc$ by:
\begin{displaymath}
\Delta E \equiv \sum_i E_i - E_{\rm beam} \hspace{0.01\columnwidth} {\rm ,}
\hspace{0.07\columnwidth}
\Mbc \equiv \sqrt{ E_{\rm beam}^2 - \big|\sum_i {\mathbf{p}}_i \big|^2 } {\rm ,}
\end{displaymath}
where $E_i$ and ${\mathbf{p}}_i$ are the energies and momenta of the $D$ decay products.  For true $D$ candidates, $\Delta E$ will be consistent with zero, and $\Mbc$ will be consistent with the $D$ mass.

We measure the branching fractions for the decays $D\to K_S^0 \pi$ by directly reconstructing the final-state particles, where the $K_S^0$ is reconstructed from $K_S^0 \to \pi^+\pi^-$.  The decay $D^+ \to \Ks\pi^+$ is measured by a separate CLEO-c analysis \cite{bib:Dhad}, but CLEO-c has not previously measured $D^0 \to \Ks\pi^0$.  In this paper, we cite the $D^+ \to \Ks\pi^+$ result and present a measurement of $D^0 \to \Ks\pi^0$.

The decays $D\to K_L^0 \pi$ have not been previously measured due to the difficulty of $K_L^0$ reconstruction.  Fortunately, the clean $D\bar{D}$ environment allows us to measure these decays without directly detecting the $K_L^0$.  Instead, we reconstruct all particles in the event except for the $\Kl$ -- that is, a tag $\bar{D}$ and a $\pi$ -- and infer the presence of a $\Kl$ from the missing four-momentum.  Our signal is a peak in the missing mass squared distribution at the $K_L^0$ mass squared.

The situation for $D^0$ decays has an added complication.
When $D^0$ and $\bar{D}^0$ are pair-produced through a virtual photon ($J^{PC}=1^{--}$),
they are in a quantum coherent state. Therefore the decays of $D^0$ and $\bar{D}^0$
are subject to interference.  This interference has no effect on the overall rate for any particular $D^0$ or $\bar{D}^0$ decay, but it does alter how often a particular $D^0$ decay occurs in combination with a particular $\bar{D}^0$ decay.  Therefore, when $D^0$ decays are measured with a reconstructed tag $\bar{D}^0$, the apparent ``branching fractions'' of the $D^0$ will vary according to the decay of the $\bar{D}^0$ \cite{bib:wsun}.  This effect is especially large for $CP$ eigenstate modes, such as $\Ks\pi^0$ and $\Kl\pi^0$.

The quantum correlation effects are shown in Table~\ref{table:QC}, where
$X$ stands for all modes combined, $f$ stands for a flavored mode, $S_+$ stands
for a $CP$-even mode, $S_-$ stands for a $CP$-odd mode, $R_{WS,f}$ is the wrong-sign decay ratio ${\mathcal B}(\bar{D} \to f)/{\mathcal B}(D \to f)$, $y$ is the $D^0$-$\bar{D}^0$ mixing parameter $y \equiv \Delta\Gamma / 2\Gamma$, and $r_fe^{-i\delta_f} \equiv \langle f | \bar{D}^0\rangle / \langle f | D^0 \rangle$.  An untagged measurement is not altered relative to a measurement using an isolated $D^0$.  However, measurements of $\Ks\pi^0$ ($CP$-odd) and $\Kl\pi^0$ ($CP$-even), tagged by a flavored $\bar{D}^0$ decay, are altered by factors of $(1+R_{WS,f} \mp r_fz_f \mp y)$.  These factors depend on the tag mode, and $z_f \equiv 2\cos\delta_f$ is generally not known.  Since $D^0 \to K_L^0 \pi^0$ must be reconstructed with a tag $\bar{D}^0$, we must determine the factor for each tag mode, $f$.  We do this by comparing tagged and untagged $D^0 \to K_S^0 \pi^0$.

\begin{table}
\centering
\caption{Untagged (vs. $X$) and tagged (vs. $f$) efficiency-corrected yields for $C=-1$ $D^0 \bar{D}^0$ events, to
leading order in the mixing parameters.  $N$ is the number of $D^0\bar{D}^0$ events, $B_i$ is the branching fraction for mode $i$ for an isolated $D^0$, and $z_f \equiv 2\cos\delta_f$.}
\label{table:QC}
\begin{tabular}{c@{~~~~}c@{~~~~}c}
\hline
\hline
      & $X$                 & $f$                                 \\
\hline
$S_+$ & $2NB_{S_+}$         & $NB_fB_{S_+}(1+R_{WS,f}+r_fz_f+y)$ \\
$S_-$ & $2NB_{S-}$          & $NB_fB_{S_-}(1+R_{WS,f}-r_fz_f-y)$ \\
$X$   & --                  & $2NB_f(1+R_{WS,f})$                 \\
\hline
\hline
\end{tabular}
\end{table}

Our procedure is the following:  We first measure the branching fraction $\mathcal B(D^0 \to K_S^0 \pi^0)$ by reconstructing this decay without tagging a $\bar{D}^0$.
Next, we measure the ``branching fraction'' for $D^0 \to K_S^0 \pi^0$, with three
different flavor tags. Each gives us
$\mathcal B(D^0 \to K_S^0 \pi^0)(1-C_f)$, where $C_f \equiv (r_fz_f+y)/(1+R_{WS,f})$. Using
$\mathcal B(D^0 \to K_S^0 \pi^0)$ from the untagged measurement, we obtain
$C_f$ for each flavor tag.
Finally, we measure the ``branching fraction'' for $D^0 \to K_L^0 \pi^0$, with the same
three flavor tags. Each gives us
$\mathcal B(D^0 \to K_L^0 \pi^0)(1+C_f)$. Using the calculated values of
$C_f$, we obtain
$\mathcal B(D^0 \to K_L^0 \pi^0)$ from each of the three tags.  These measurements are then averaged for the final result.

We first measure $D^0 \to \Ks\pi^0$ without searching for a tag $\bar{D}^0$.  Candidates for $D^0 \to K_S^0 \pi^0$ are formed by combining
a $K_S^0$, reconstructed by a pair of charged tracks through the decay
$K_S^0 \to \pi^+ \pi^-$, and a $\pi^0$, reconstructed from a pair of photons detected in the
CsI calorimeter.  The invariant mass of $\Ks$ and $\pi^0$ candidates is required to be consistent with the known mass, and $\pi^0$ candidates are then constrained to the known mass.

Both beam-constrained mass and $\Delta E$ are required to be within
3 standard deviations of the nominal value.  If there are multiple candidates in one event, we accept only the one whose beam-constrained mass is closest to the nominal $D^0$ mass.  Two sideband subtractions are used to remove
background.  First, a $\Delta E$ sideband subtraction is used to remove the continuum
and combinatoric background (a 13\% effect).  Then, a $K_S^0$ mass sideband subtraction is used to remove the background from $D^0 \to \pi^+ \pi^- \pi^0$ events in which $M(\pi^+ \pi^-)$ happens to be within the $K_S^0$ mass window (a 4\% effect).  The
resulting yield is 7487$\pm$101 events.  This yield is divided by the detection efficiency, 29.3\%, to determine the number of $D^0 \to K_S^0 \pi^0$ events produced.  The efficiency is determined from Monte Carlo simulation, with a correction for $\pi^0$ detection efficiency; this correction is determined by comparing $\pi^0$ efficiencies measured in data and in our simulation.
Finally, we use the total number of $D^0\bar{D}^0$ events in our sample, $1.031 \times 10^6$ (from a separate CLEO-c analysis~\cite{bib:Dhad}).  Dividing the efficiency-corrected yield by twice this number gives the branching fraction.

Systematic uncertainties considered include those from:  the $\Delta E$ cut
($\pm$0.5\%), the $\Delta E$ sideband subtraction ($\pm$0.8\%), tracking efficiency
($\pm$0.6\%), $\Ks$ detection efficiency ($\pm$1.8\%), the $\Ks$ sideband subtraction
($\pm$0.3\%), and the number of $D^0\bar{D}^0$ events ($\pm$1.4\%).  These total $\pm$2.5\%.  The largest uncertainty is due to $\pi^0$ reconstruction efficiency ($\pm$3.8\%).  Although this uncertainty is large, it cancels in the computation of quantum correlation factors and in the comparison of the $D^0 \to \Ks\pi^0$ and $D^0 \to \Kl\pi^0$ branching fractions.  Therefore we keep it separate from the other uncertainties.

We find a branching fraction
${\mathcal B}(D^0 \to K_S^0 \pi^0) = (1.240 \pm 0.017 \pm 0.031 \pm 0.047)\%$, where the last uncertainty is from the $\pi^0$ efficiency.

Having determined ${\mathcal B}(D^0 \to K_S^0 \pi^0)$, we now measure this decay with three different tag modes to obtain the quantum correlation factors $(1-C_f)$.  The three tag modes we use are $\bar{D}^0 \to K^+ \pi^-$, $\bar{D}^0 \to K^+ \pi^- \pi^0$, and $\bar{D}^0 \to K^+ \pi^- \pi^- \pi^+$.  The tag $\bar{D}^0$ is required to
be within 3 standard deviations of the nominal values of $\Delta E$ and $\Mbc$.  We select at most one candidate per flavor per tag mode; when multiple candidates pass our requirements, we keep the one with $\Mbc$ closest to the nominal $D^0$ mass.  We remove fake tag $\bar{D}^0$ candidates by subtracting the $\Delta E$ sideband of the tag.

In the tagged sample, we reconstruct $D^0 \to K_S^0 \pi^0$ in the same way as
in the untagged case.  To remove fake $D^0 \to \Ks\pi^0$ candidates, we subtract a $\Ks$ mass sideband.  (No $\Delta E$ sideband subtraction is necessary since, with a tag, the $\Ks\pi^0$ signal is essentially free of combinatoric background.)

Although to first order the efficiency of reconstructing the $\bar{D}^0$ tag
cancels in the branching fraction
calculation, simulations indicate a slightly larger
efficiency for $\bar{D}^0$ reconstruction
when the signal $D^0$ decays to $\Ks\pi^0$. This bias stems from the
lower-than-average multiplicity
of particles in $D^0 \to \Ks\pi^0$ events.  We obtain correction
factors for these small biases from Monte Carlo studies.

With the efficiencies from Monte Carlo simulations and the yields in signal and sideband regions, we compute the branching
fractions, times quantum correlation factors, in Table~\ref{table:Ks-DT}.

\begin{table}
\centering
\caption{Efficiencies, yields, and results for tagged $D^0 \to K_S^0 \pi^0$ study.  No systematic uncertainties are included in the quoted results.}
\label{table:Ks-DT}
\begin{tabular}{llll}
\hline
\hline
Tag mode                           & $K^+\pi^-$ & $K^+\pi^-\pi^0$ & $K^+\pi^-\pi^-\pi^+$ \\
\hline
Efficiency                         & 31.74\%            & 31.29\%                 & 29.97\% \\
Tag yield - raw                    & 48095              & 67576                   & 75113   \\
Sideband subtracted                & 47440              & 63913                   & 71040   \\
Signal yield - raw                 & 172                & 248                     & 276     \\
Sideband subtracted                & 155                & 203                     & 256     \\
Tag bias correction                & 1.000              & 1.014                   & 1.033   \\
\hline
$\mathcal B(K_S^0 \pi^0)(1-C_f)$ (\%) & 1.03$\pm$0.09   & 1.00$\pm$0.09           & 1.16$\pm$0.08 \\
\hline
\hline
\end{tabular}
\end{table}

The systematic uncertainties are similar to those in the untagged measurement.  Track, $K_S^0$, and $\pi^0$
reconstruction uncertainties are the same, and they
will cancel in the ratio of the tagged and untagged results.  The only systematic uncertainties from the tag $\bar{D}^0$ are for the $\Delta E$ sideband subtraction and the tag bias correction factor; any other discrepancies in the Monte Carlo simulation would have the same effect on the tag and signal yields.

Finally, we divide these results by ${\mathcal B}(D^0 \to K_S^0 \pi^0)$, from the untagged measurement, to obtain the three quantum correlation factors $(1-C_f)$, where $f$ represents the tag mode.

We measure the $D \to \Kl\pi$ branching fractions with a missing mass technique.  We reconstruct the tag $\bar{D}$ in 3 $\bar{D}^0$ modes and 6 $D^-$ modes, and we combine it with a $\pi^0$ or $\pi^+$ to form missing mass squared: $\MMsq \equiv (p_{\rm event} - p_{\bar{D}} - p_\pi)^2$.  To improve resolution, the tag $\bar{D}$ is constrained to have the expected three-momentum magnitude.  The $D \to \Kl\pi$ signal is a peak in $\MMsq$ at the $K_L^0$ mass squared ($\sim$ 0.25 $\rm{GeV}^2$).

To remove $D \to \Ks\pi$ events, as well as other backgrounds, we require that the event contain no extra tracks or $\pi^0$'s beyond those used in the tag $\bar{D}$ and the $\pi$.  This veto removes about 90\% of $D \to \Ks\pi$ events and a few percent of $D \to \Kl\pi$ events.  For $D^0 \to K_L^0 \pi^0$ only, we also remove an event if it contains an extra $\eta \to \gamma\gamma$.  This removes much of the $D^0 \to \eta \pi^0$ background.  To determine systematic uncertainties from the appearance of fake extra particles in signal events, we compare how often they appear in data and in our simulation, using events in which both $D$ and $\bar{D}$ were fully reconstructed.

As in the tagged $D^0 \to K_S^0 \pi^0$ study, the tag $\bar{D}$ reconstruction efficiency is higher when the $D$ decays to $\Kl\pi$; therefore we apply correction factors determined from Monte Carlo simulations.  The efficiency for observing $D \to \Kl \pi$, given that the tag was found, is also determined in these simulations.  It is essentially the efficiency for finding the $\pi$ without any fake extra particles.

\begin{figure}
\centering
\includegraphics[width=1.0\columnwidth]{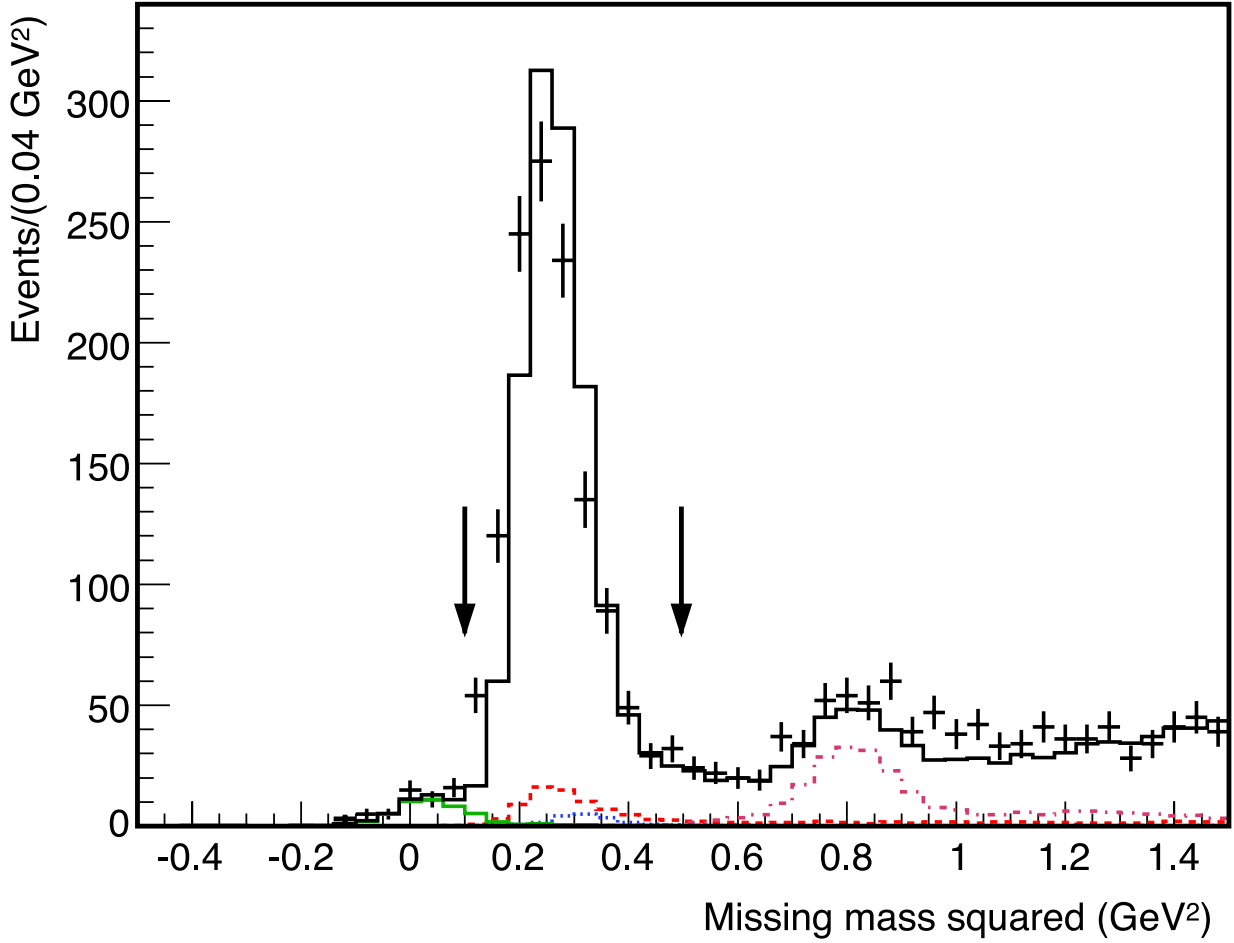}
\caption{Missing mass squared distribution, with all tag modes combined, for $D^0 \to X \pi^0$, after removing events with extra tracks, $\pi^0$'s, or $\eta$'s.  The points with error bars are data, and the solid line is a Monte Carlo simulation.  The dashed, colored lines represent simulations of the peaking backgrounds $D^0 \to \pi^0\pi^0$, $\Ks\pi^0$, $\eta\pi^0$, and $K^{*0}\pi^0$.  The difference in the peak position is due to a minor discrepancy in our calorimeter simulation at large photon energies; the signal region, marked with arrows, encompasses the peak in both distributions.}
\label{fig:Klpi0}
\end{figure}

For the $D^0 \to K_L^0 \pi^0$ branching fraction measurement,
the same three $\bar{D}^0$ decay modes are selected with the same requirements as in the tagged
$D^0 \to K_S^0 \pi^0$ study (except for a minor difference in the order of applying cuts
for the $K^+ \pi^- \pi^0$ tag, which results in a slight difference in number of tags).  Combining these $\bar{D}^0$ candidates with $\pi^0$ candidates and rejecting events with extra tracks, $\pi^0$'s, or $\eta$'s, we obtain the $\MMsq$ plot shown in Fig.~\ref{fig:Klpi0}.

A number of backgrounds slip through our extra track, $\pi^0$, and $\eta$ vetoes and appear in the $\MMsq$ plot.  The modes $\Ks\pi^0$ and $\eta\pi^0$ appear as peaks at essentially the same location as $\Kl\pi^0$; $\pi^0\pi^0$ peaks at $\MMsq \approx 0.0~{\rm GeV}^2$; and $K^{*0}\pi^0$ peaks at 0.8 GeV$^2$.  Monte Carlo simulations of these backgrounds are shown in Fig.~\ref{fig:Klpi0}.  Other, lesser backgrounds also appear to the right of the $\Kl\pi^0$ peak.

To determine the signal and estimate the background, we define a $\MMsq$ signal region $0.1$ to $0.5 ~\rm GeV^2$, as well as low and high sidebands: $-0.1$ to $0.1 ~\rm GeV^2$ and $0.8$ to $1.2 ~\rm GeV^2$.  The backgrounds are split into three groups: $D^0 \to K_S^0 \pi^0$ and $D^0 \to \eta \pi^0$; $D^0 \to \pi^0 \pi^0$; and all other backgrounds.  For $D^0 \to K_S^0 \pi^0$ and $D^0 \to \eta \pi^0$, we use Monte Carlo simulation to determine efficiencies for the background
subtraction. For $D^0 \to \pi^0 \pi^0$, we scale the contribution to the signal region according to the yield in the low sideband.  For the sum of all other backgrounds, we follow the same procedure with the high sideband.  In total, about 10\% of the events in the signal region are background, with half coming from $\Ks\pi^0$, 1/10 from each of $\eta\pi^0$ and $\pi^0\pi^0$, and 3/10 from various other decays.

\begin{table}
\centering
\caption{Efficiencies, yields, and results for $D^0 \to K_L^0 \pi^0$.  No systematic uncertainties are included in the quoted results.}
\label{table:all}
\begin{tabular}{llll}
\hline
\hline
Tag mode                           & $K^+\pi^-$ & $K^+\pi^-\pi^0$ & $K^+\pi^-\pi^-\pi^+$ \\
\hline
Efficiency                         & 55.21\%            & 52.72\%                 & 49.88\% \\
Tag yield - raw                    & 48095              & 68000                   & 75113   \\
Sideband subtracted                & 47440              & 64280                   & 71040    \\
Signal yield - raw                 & 367.0              & 414.5                   & 466.5   \\
Background subtracted              & 334.8              & 363.1                   & 418.0   \\
Tag bias correction                & 1.000              & 1.037                   & 1.057   \\
\hline
$\mathcal B(K_L^0 \pi^0)(1+C_f)$ (\%) & 1.28$\pm$0.08   & 1.03$\pm$0.06           & 1.12$\pm$0.06 \\
\hline
\hline
\end{tabular}
\end{table}

After subtracting all the backgrounds, we obtain the yields and compute branching
fractions, times quantum correlation factors, in Table~\ref{table:all}.

Systematic uncertainties come from the effect on signal efficiency of the veto on extra tracks ($\pm$0.3\%), the veto on extra $\pi^0$'s ($\pm$1.6\%), the veto on $\eta$'s ($\pm$0.5\%), and the uncertainty in the location and width of the signal peak ($\pm$1.4\%).  Other uncertainties come from the background estimate ($\pm$1.0\%), $\Delta E$ sideband subtraction ($\pm$0.5\%), and the tag bias correction factor ($\pm$0.2\%).  These total $\pm$2.5\%.  As in $D^0 \to \Ks\pi^0$, $\pi^0$ efficiency ($\pm$3.8\%) is the largest systematic uncertainty; it cancels in the comparison of $D^0 \to \Ks\pi^0$ and $D^0 \to \Kl\pi^0$.

We have determined $\mathcal B(D^0 \to K_L^0 \pi^0)(1+C_f)$ for three different flavor tags $f$.
Using the values of $C_f$ determined from the $D^0 \to K_S^0 \pi^0$ measurements, we
calculate $\mathcal B(D^0 \to K_L^0 \pi^0)$ for each tag mode.  Finally, we average the results and find $\mathcal B(D^0 \to K_L^0 \pi^0) = (0.998 \pm 0.049 \pm 0.030 \pm 0.038)\%$, where the last uncertainty is from the $\pi^0$ efficiency.

The analysis of $D^+ \to \Kl\pi^+$ is similar to $D^0 \to \Kl\pi^0$, though there are a few differences.  Since we reconstruct a $\pi^+$ instead of a $\pi^0$, the $\MMsq$ resolution is better.  Also, we do not need to correct for quantum correlation.  The most significant difference in procedure is that we perform a likelihood fit for the signal and background yields instead of counting events in a signal region.

We reconstruct tag $D^-$'s in six decay modes: $D^- \to K^+\pi^- \pi^-$, $K^+ \pi^- \pi^- \pi^0$, $\Ks \pi^-$, $\Ks \pi^- \pi^0$, $\Ks \pi^- \pi^- \pi^+$, and $K^+ K^- \pi^-$.  As before, candidates must have $\Delta E$ consistent with zero.  We select one candidate per charge per mode based on the best value of $\Delta E$.  We fit the $\Mbc$ distribution for each mode to determine the number of tags, and then pass all candidates with $\Mbc$ near the peak to be combined with $\pi^+$ candidates.

\begin{figure}
\begin{center}
 \includegraphics[width=1.0\columnwidth]{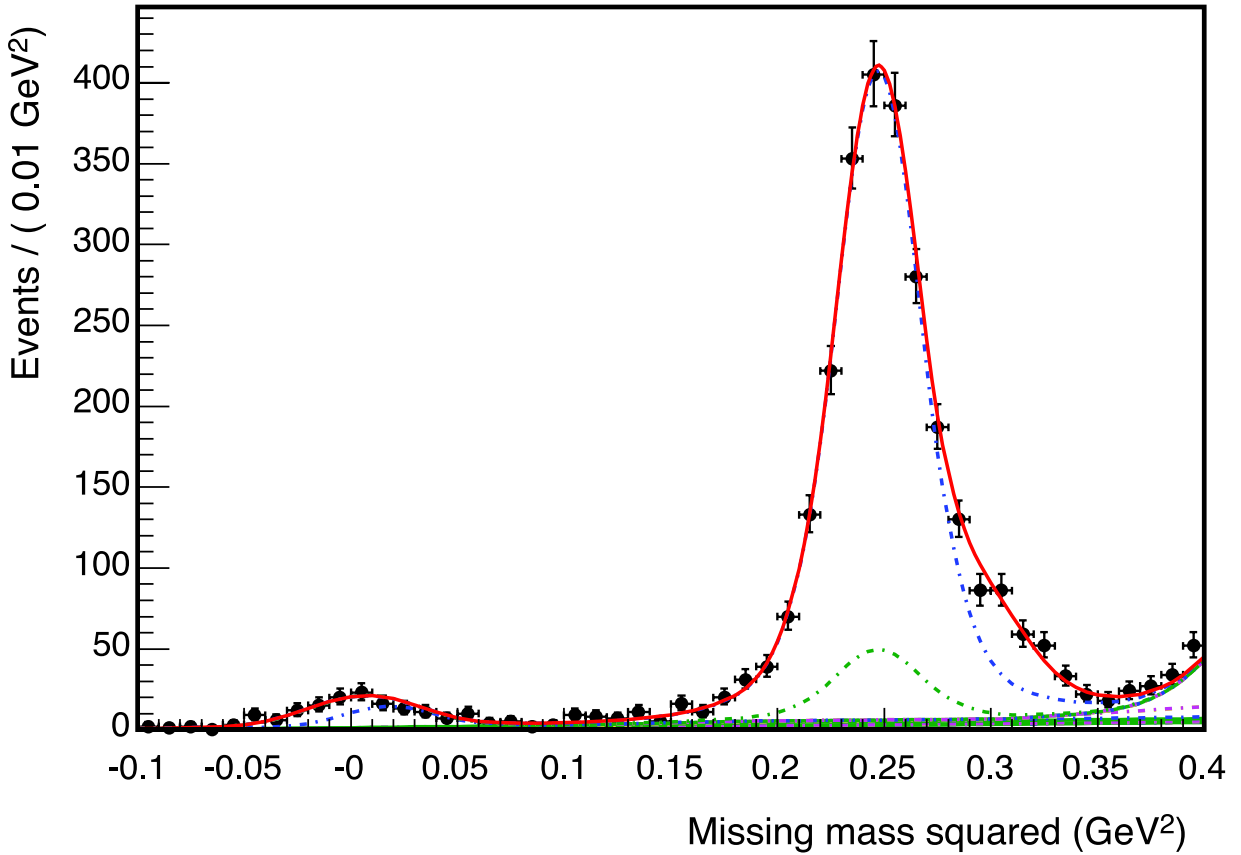}
\end{center}
\caption{Missing mass squared distribution, with all tag modes combined, for $D^+ \to X \pi^+$, after removing events with extra tracks or $\pi^0$'s.  The solid line shows a fit for the $D^+ \to \Kl\pi^+$ yield.  The many dashed lines represent the various components of the fit, added cumulatively.  The small peak under the signal is the contribution of $D^+ \to \Ks\pi^+$ events that are not removed by the extra track and $\pi^0$ vetoes.}
\label{fig:KlpiMMsqFit}
\end{figure}

The $\MMsq$ distribution, with all tag modes added together, is shown in Fig.~\ref{fig:KlpiMMsqFit}.  The lines show a fit used to determine the signal yield.  The most prominent feature is the signal peak at the $\Kl$ mass squared ($\sim$0.25 $\rm{GeV}^2$).  A number of backgrounds are also present.  First, fake $D^-$ candidates produce a background which is estimated from an $\Mbc$ sideband.  All of the other backgrounds come from other $D^+$ decays.  The largest of these are $D^+ \to \Ks\pi^+$ (dashed, green peak under the signal), $\eta\pi^+$ (shoulder on the right-side tail of the signal), $\pi^0\pi^+$ and $\mu^+ \nu_\mu$ (peak on the left of the plot), $\bar{K}^0 \pi^+ \pi^0$, and $\pi^+\pi^0\pi^0$.  The shapes and efficiencies of these backgrounds are determined from Monte Carlo simulations.  The yields of the signal peak and the $\eta\pi^+$, $\pi^0\pi^+$, and $\mu^+ \nu_\mu$ backgrounds are allowed to vary in the fit; all other yields are fixed based on the efficiencies.

Although Fig.~\ref{fig:KlpiMMsqFit} shows all tag modes together, we actually fit each tag mode separately.  We calculate a branching fraction from each tag mode using the tag bias correction factor, efficiency, tag $D^-$ yield, and signal $D^+ \to \Kl\pi^+$ yield for that mode.  The tag bias correction varies from 1.005 (for $K^+\pi^-\pi^-$) to 1.047 (for $\Ks\pi^-\pi^-\pi^+$).  The efficiency averages to 81.6\%, and depends little on tag
mode.  There are a total of $165 \times 10^3$ tags, and a total $D^+ \to \Kl\pi^+$ yield of
2023$\pm$54 events.  The values of the branching fraction calculated from each tag mode are averaged to produce the final result.

Systematic uncertainties include those from: pion reconstruction efficiency ($\pm$0.3\%) and particle identification ($\pm$0.25\%), tag bias correction factor ($\pm$0.2\%), charge of the tag $D$ ($\pm$0.5\%), extra track and extra $\pi^0$ vetoes ($\pm$1.1\%), signal peak shape ($\pm$0.7\%), signal peak width ($\pm$1.6\%), contribution of fake $D^-$ tags ($\pm$0.4\%), and $\MMsq$ background yields ($\pm$0.8\% from statistical uncertainty in $\Ks\pi^+$ background, $\pm$0.3\% from $\BF(D^+\to \Ks\pi^+)$, and $\pm$0.5\% from all other backgrounds).  The total systematic uncertainty is $\pm$2.4\%.

We find a branching fraction ${\mathcal B}(D^+ \to \Kl  \pi^+)$ = (1.460 $\pm$ 0.040 $\pm$ 0.035 $\pm$ 0.005)\%.  The final uncertainty is due to the input value of ${\mathcal B}(D^+ \to \Ks \pi^+)$.

To compare $D \to \Ks\pi$ and $D \to \Kl\pi$, we compute the asymmetries
\begin{displaymath}
R(D) \equiv \frac{ \BF(D \to \Ks \pi) - \BF(D \to \Kl \pi) }{ \BF(D \to \Ks \pi) + \BF(D \to \Kl \pi) } {\rm .}
\end{displaymath}
The $D^0$ asymmetry (in which the systematic uncertainty for $\pi^0$ efficiency cancels) is $R(D^0) = 0.108 \pm 0.025 \pm 0.024$.  Using $\BF(D^+ \to \Ks\pi^+) = (1.526 \pm 0.022 \pm 0.038)\%$ \cite{bib:Dhad}, the $D^+$ asymmetry is $R(D^+) = 0.022 \pm 0.016 \pm 0.018$.

The asymmetry between $D^0 \to \Ks\pi^0$ and $D^0 \to \Kl\pi^0$ is consistent with SU(3) symmetry, and in particular the U-spin subgroup of SU(3).  U-spin predicts $A(D^0 \to K^0\pi^0)/A(D^0 \to \bar{K}^0\pi^0) = -\tan^2\theta_C$, where $\theta_C$ is the Cabibbo angle.  This prediction is relatively insensitive to SU(3) breaking \cite{bib:RosnerUspin}.  The amplitude ratio can also be predicted from diagrams for these two processes; both have spectator and exchange diagrams which differ only by a factor of $-\tan^2\theta_C$.  However derived, the amplitude ratio implies that the asymmetry is $R(D^0) = 2 \tan^2\theta_C$.  Using $\tan\theta_C = 0.233 \pm 0.001$ \cite{bib:PDG2006}, we calculate $R(D^0) = 0.109 \pm 0.001$, in good agreement with our measurement.

There is no corresponding U-spin argument for the $D^+$ decays, so no simple prediction is possible.  Diagrams for the Cabibbo-favored and doubly-suppressed decays are different.  Both internal and external spectator diagrams contribute to $D^+ \to \bar{K}^0\pi^+$, while $D^+ \to K^0 \pi^+$ has internal spectator and annihilation diagrams.  Approximate predictions are, however, possible under certain assumptions.  One analysis \cite{bib:Gao}, based on flavor SU(3) with an estimate of symmetry-breaking
effects, finds $R(D^+) \approx 0.04$, consistent with our measurement.
This analysis also points out that the small asymmetry found
for $D^+$ decays can be interpreted as a large strong phase
between two contributing amplitudes
in the case of $D^+$ decays, while the larger asymmetry in the
$D^0$ decays is consistent with a small strong phase.

We gratefully acknowledge the effort of the CESR staff
in providing us with excellent luminosity and running conditions.
This work was supported by
the A.P.~Sloan Foundation,
the National Science Foundation,
the U.S. Department of Energy, and
the Natural Sciences and Engineering Research Council of Canada.


\begin{thebibliography}{99}

\bibitem{bigi}   I.I.~Bigi and H.~Yamamoto, Physics Letters B \textbf{349}, 363 (1995).
\bibitem{bib:cleo_1} Y.~Kubota {\it et al.} (CLEO Collaboration), Nucl. Instrum. Methods Phys. Res., Sect. A \textbf{320}, 66 (1992).
\bibitem{bib:cleo_2} D.~Peterson {\it et al.}, Nucl. Instrum. Methods Phys. Res., Sect. A \textbf{478}, 142 (2002).
\bibitem{bib:cleo_3} M.~Artuso {\it et al.}, Nucl. Instrum. Methods Phys. Res., Sect. A \textbf{554}, 147 (2005).
\bibitem{bib:MarkIII} J.~Adler {\it et al.} (Mark III Collaboration), Phys. Rev. Lett. \textbf{62}, 1821 (1989).
\bibitem{bib:Dhad} S.~Dobbs {\it et al.} (CLEO Collaboration), Phys. Rev. D. \textbf{76}, 112001 (2007).
\bibitem{bib:wsun} D.M.~Asner and W.M.~Sun, Phys. Rev. D \textbf{73}, 034024 (2006); \textbf{77}, 019901(E) (2008).
\bibitem{bib:RosnerUspin} J.L.~Rosner, Phys. Rev. D {\bf 74}, 057502 (2006).
\bibitem{bib:PDG2006}   W.-M.~Yao {\it et al.}, Journal of Physics G \textbf{33}, 1 (2006).
\bibitem{bib:Gao} D.-N.~Gao, Physics Letters B \textbf{645}, 59 (2007).
\end{thebibliography}
\end{document}